# Strain and band gap engineering in GeSn alloys via P doping

S. Prucnal[1,a)], Y. Berencén[1], M. Wang[1], J. Grenzer[1], M. Voelskow[1], R. Hübner[1], Y. Yamamoto[3], A. Scheit[3], F. Bärwolf[3], V. Zviagin[5], R. Schmidt-Grund[5], M. Grundmann[5], J. Żuk[2], M. Turek[2], A. Droździel[2], K. Pyszniak[2], R. Kudrawiec[6], M. P. Polak[6], L. Rebohle[1], W. Skorupa[1], M. Helm[1,4] and S. Zhou[1]

[1] *Institute of Ion Beam Physics and Materials Research, Helmholtz-Zentrum Dresden-Rossendorf, P.O. Box 510119, 01314 Dresden, Germany*

[2] *Maria Curie-Sklodowska University, Pl. M. Curie-Sklodowskiej 1, 20-035 Lublin, Poland*

[3] *IHP, Im Technologiepark 25, 15236 Frankfurt (Oder), Germany*

[4] *Center for Advancing Electronics Dresden (cfaed), Technische Universität Dresden, 01062 Dresden, Germany*

[5] *Felix-Bloch-Institut für Festkörperphysik, Universität Leipzig Linnéstr. 5, 04103 Leipzig, Germany*

[6] *Faculty of Fundamental Problems of Technology, Wroclaw University of Science and Technology, Wybrzeże Wyspiańskiego 27, 50-370 Wrocław, Poland*

a) Corresponding author: s.prucnal@hzdr.de

## Abstract

Ge with a quasi-direct band gap can be realized by strain engineering, alloying with Sn or ultra-high n-type doping. In this paper, we use all three approaches together - strain engineering, Sn alloying and n-type doping to fabricate direct band gap GeSn alloys. The heavily-doped n-type GeSn was realized using a CMOS-compatible non-equilibrium material processing. P is used to form a highly-doped n-type GeSn layers and to modify the lattice parameter of GeSn:P alloys. The strain engineering in heavily P-doped GeSn films is confirmed by X-ray diffraction and micro-Raman spectroscopy. The change of the band gap in GeSn:P alloy as a function of P concentration is theoretically predicted using density functional theory and experimentally verified by near-infrared spectroscopic ellipsometry. According to the shift of the absorption edge it is shown that for an electron concentration above $1\times10^{20}$ cm$^{-3}$ the band gap renormalization is partially compensated by the Burstein–Moss effect. These results indicate that Ge-based materials have a large potential for the near-infrared optoelectronic devices, fully compatible with CMOS technology.

# 1. Introduction

Ge is one of the most promising materials which can boost the performance of nanoelectronic devices even further. Due to the chemical and physical similarity to Si, Ge can be integrated into CMOS technology. Intrinsic bulk Ge is an indirect band-gap material, but it can be converted to a quasi-direct band gap material due to the small separation between the Γ and L valley (136 meV) [1]. It was shown that the direct band gap of Ge can be realized by tensile strain engineering, alloying with Sn or ultra-high n-type doping [2-5]. By using only one of these methods it is extremely difficult to achieve a direct band gap. The biaxial tensile strain required for the direct band gap formation in Ge has to be above 1.9 % [6]. This is achieved by growing Ge on foreign substrates with a lattice parameter much larger than that of Ge e.g. on InGaAs [7]. Unfortunately, this approach requires the use of GaAs substrates and processes which are not compatible with CMOS technology.

Kimerling's group has grown relaxed Ge on a Si substrate using high-temperature annealing and fast cooling [5]. Due to the different expansion coefficient between Si and Ge the created layer has a tensile strain at a level of 0.25%. The predicted maximal biaxial tensile strain which can be obtained by this method (0.35%) is far below the required 1.9 % to form a direct band gap in Ge. Nevertheless, the remaining difference between the Γ and L valley (115 meV) can be compensated by n-type doping. The minimum electron concentration needed to completely fill the L valley and to start the population of the Γ valley is in the range of $8\times10^{19}$ cm$^{-3}$. The main advantage of this method is the fact that the main optical emission under such conditions is at about 1550 nm [5] which matches perfectly with the optical window of quartz fibers, but attaining such doping level remains challenging.

Recently, much more effort has been taken on the fabrication of direct band gap GeSn alloys. It was shown that GeSn with a Sn concentrations higher than 8 % can be converted to a direct band gap semiconductor, which shows optically-driven lasing at low temperatures [8, 9]. The compressive strain formed in GeSn due to the lattice mismatch between GeSn and Ge increases the direct gap which in-turn requires even higher concentrations of Sn to achieve a direct gap.

In this paper, we report on a combined approach towards direct band gap GeSn by co-doping with P which in-turn does not only compensate the biaxial compressive strain in GeSn, but also introduces ultra-high n-type doping. The ultra-doped n-type GeSn layer-on-Si is fabricated by ion implantation of Sn and P into Ge-on-Si wafers followed by non-equilibrium ms-range rear-side flash-lamp annealing (r-FLA). Recrystallization of the implanted layer and the electrical activation of P induce explosive solid-phase epitaxy (ESPE) regrowth [10, 11]. According to XRD data the lattice parameter of $Ge_{0.97}Sn_{0.03}$:P alloy in comparison to the relaxed $Ge_{0.97}Sn_{0.03}$ binary alloy is reduced by 0.42% ($a_{Ge0.97Sn0.03:P}$=5.659 Å vs. $a_{Ge0.97Sn0.03}$=5.683 Å). The band gap change of the fabricated $Ge_{0.97}Sn_{0.03}$:P alloys cannot be explained just by either the alloy formation or the band gap renormalization. The influence of the reduction of the lattice parameter on the band gap change together with the Burstein-Moss effect has to be considered. The fabricated GeSn:P alloys are n-type with an electron concentration as high as $1\times10^{20}cm^{-3}$.

### 2. Experimental section

A 475 nm thick undoped Ge layer was epitaxially grown on a Si (100) substrate using reduced-pressure chemical vapor deposition at IHP Frankfurt/Oder. Prior to ion implantation, a 30 nm thick $SiO_2$ layer was deposited by plasma enhanced chemical vapor deposition. $SiO_2$ was used to prevent the Ge surface from roughening and sputtering during ion implantation [12, 13]. Subsequently, samples were implanted with Sn and P (see Table 1 for details). After ion implantation, an ultra-short non-equilibrium thermal processing was applied. Prior to this step, samples were preheated to a temperature of 180 $^o$C for 30 s in nitrogen ambient. Samples were subsequently annealed using r-FLA for 3 ms at an energy density of 61.5 $Jcm^{-2}$ deposited onto the rear-side of sample. The r-FLA process is described in detail elsewhere [10]. During recent years, flash lamp annealing (FLA) has been used as an annealing method in the ms-range for many purposes, especially driven by the needs of advanced semiconductor processing, see Refs. 14-16 for recent reviews, and more specific, also those of Ge-based materials research: shallow junctions [17], nanoclusters [18], advanced processing [10, 19] etc.

The incorporation efficiency of Sn into Ge during explosive recrystallization was investigated using random and channeling Rutherford backscattering spectrometry (RBS/R and RBS/C). RBS was performed using the 1.7 MeV $He^+$ beam of the Rossendorf van de Graff accelerator. P and Sn concentrations versus depth profiles were measured by a dynamic secondary ion mass spectrometer (SIMS) with 1 keV $Cs^+$ and $O^+$ sputter beams, respectively. Beams of $Cs^+$ or $O^+$ ions were rastered over a surface area of 300×300 $\mu m^2$ and secondary ions were collected from the central part of the sputtered crater. Crater depths were measured with a Dektak 8 stylus profilometer, and a constant erosion rate was assumed when converting the sputtering time to the sample depth.

Table 1 Ion implantation parameters together with electron concentration and electron mobility obtained by Hall Effect measurements at 2.5 K and 300 K, respectively.

| Sample name | Sn fluence at 250 keV ($cm^{-2}$) | P fluence at 80 keV ($cm^{-2}$) | Carrier concentration ($cm^{-3}$) | | Carrier mobility at 300 K ($cm^2/Vs$) |
|---|---|---|---|---|---|
| | | | 2.5 K | 300 K | |
| GeP2 | - | $6.0 \times 10^{15}$ | $9.1 \times 10^{19}$ | $9.6 \times 10^{19}$ | 162 |
| GeSn | $1.0 \times 10^{16}$ | - | | | |
| GeSnP1 | $1.0 \times 10^{16}$ | $3.0 \times 10^{15}$ | $7.0 \times 10^{19}$ | $7.2 \times 10^{19}$ | 140 |
| GeSnP2 | $1.0 \times 10^{16}$ | $6.0 \times 10^{15}$ | $8.5 \times 10^{19}$ | $8.8 \times 10^{19}$ | 164 |
| GeSnP3 | $1.0 \times 10^{16}$ | $9.0 \times 10^{15}$ | $9.5 \times 10^{19}$ | $10.5 \times 10^{19}$ | 175 |

To investigate the microstructure properties of the implanted Ge layer, cross-sectional bright-field and high-resolution transmission electron microscopy (TEM) investigations were performed with a Titan 80-300 (FEI) microscope operated at an accelerating voltage of 300 kV. High-angle annular dark-field scanning transmission electron microscopy (HAADF-STEM) imaging and element mapping based on energy-dispersive X-ray spectroscopy (EDXS) were performed at 200 kV with a Talos F200X microscope equipped with a Super-X EDXS detector system (FEI). Prior to TEM analysis, the specimen mounted in a high-visibility low-background holder was placed for 10 s into a Model 1020 Plasma Cleaner (Fischione) to remove contaminations.

The strain evolution in both the implanted and the annealed samples was investigated using micro-Raman spectroscopy and X-ray diffraction (XRD). XRD was performed by an Empyrean Panalytical diffractometer with a Cu-target source. The setup is equipped with a Göbel mirror and an asymmetric monochromator to enhance the brilliance and monochromaticity.

The band-gap change in GeSnP ternary alloys was determined by means of spectroscopic ellipsometry. The thin film complex dielectric function (DF) was obtained using variable angle spectroscopic ellipsometry (VASE) in a polarizer-compensator-sample-analyzer configuration from J.A. Woolam. Measurements were conducted at room temperature and in the spectral range from 0.5 to 4.0 eV. The ellipsometry spectra were analyzed by a model consisting of a layer stack as follows: (from the bottom) the substrate, the GeSn and GeSnP film layer and a surface layer with a 50% mix of the DF of the top-most film and air. The optical constants of the modified thin layers were determined numerically by using a Kramers-Kronig consistent B-Spline approximation with a knot number choice approach described elsewhere [20]. The optical layer thickness showed a general agreement with the RBS and SIMS measurements. The obtained fit agreed well with the experimental data with a mean square error (MSE) of less than 2.2. The reflected light was detected by a Si and InGaAs diode in the UV-VIS and NIR spectral range, respectively.

The concentration of carriers in the implanted and annealed samples was estimated from temperature-dependent Hall effect measurements in a van der Pauw configuration. The thickness of the doped layer was extracted from SIMS measurements.

The band structure and position of the Fermi energy in P-doped GeSn alloys were calculated using the full potential linearized augmented plane wave (FP-LAPW) WIEN2k code in a supercell (SC) approach with the modified Becke‑Johnson local density approximation (MBJLDA) and the spectral weight approach used for the band gap correction and for band unfolding, respectively.

### 3. Results and discussion

### 3.1 Microstructural properties

The redistribution of the implanted elements and their lattice location in the Ge film were investigated using RBS performed in random and channeling direction on both as-implanted and annealed samples (see Fig.1a). The RBS random spectrum obtained from sample GeSnP2 exhibits three distinct regions corresponding to Si with the detection energies below 690 keV (the Si signal from the substrate), to Ge – registered in the energy range of 1030 and 1380 keV (475 nm thick Ge on the Si substrate), and to Sn observed between 1400 and 1500 keV. In the as-implanted stage, the RBS/C spectrum can be used to calculate the concentration of Sn in Ge and to estimate the thickness of the implanted region. Unfortunately, P was not resolvable in our RBS spectra because P is much lighter than the matrix elements and its signal overlaps with the Ge signal. The thickness of the implanted layer can be calculated from the Ge signal in the RBS/C spectrum. Due to the end-of-range defects and the large implantation fluence of P, the total thickness of the almost amorphous Ge layer formed during ion implantation is 230 nm compared to the assumed thickness of 200 nm of the P-doped layer as calculated by SRIM [21]. Considering the three P fluences and the thickness of the implanted layer, the deduced P peak concentrations are $1.5\times10^{20}$ (GeSnP1), $3.0 \times 10^{20}$ (GeSnP2) and $4.5 \times 10^{20}$ (GeSnP3) cm$^{-3}$, respectively. The Sn-doped layer is about 150 nm thick and the Sn peak concentration calculated from the RBS/R spectrum is 3.0 % in all the investigated samples. According to the RBS/R and SIMS measurements, the distribution of Sn in the annealed samples remains without changes compared to the as-implanted sample confirming that there is no remarkable diffusion of Sn during ms-range FLA (see inset Fig. 1a and b).

The RBS/C investigation performed on the annealed samples provides information about the recrystallization efficiency of the implanted layer and the lattice location of the implanted elements. In the ideal crystal, close to the surface region (first 100 nm), the yield of the RBS/C spectrum should be below 10 % of that obtained under random conditions. In the FLA treated sample the RBS/C signal from Sn is remarkably visible, which means that Sn is fully incorporated into the Ge lattice. The Sn content in the fabricated GeSn:P alloy is more than five times greater than the equilibrium solid solubility limit of Sn in Ge.

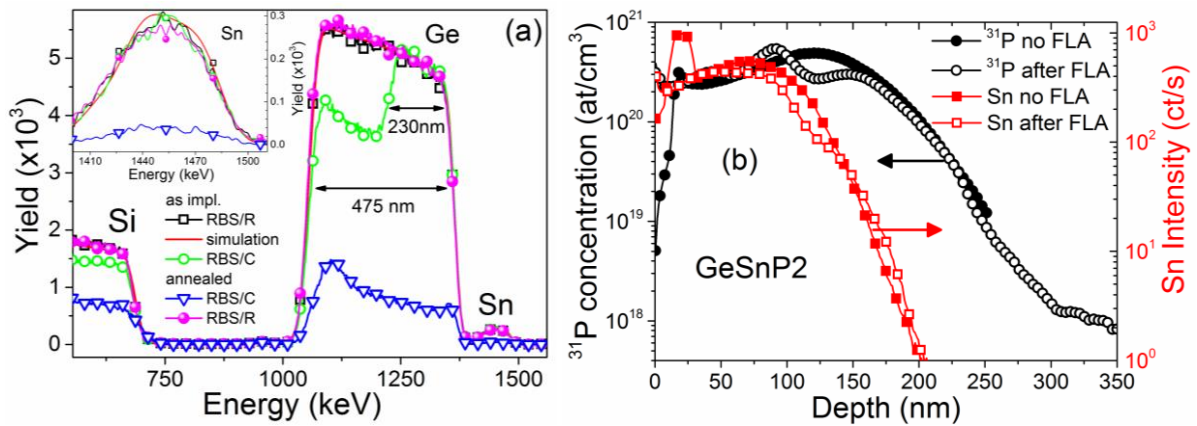

Figure 1. (a) RBS random and channeling spectra obtained from the as-implanted and the annealed GeSnP2 samples. The inset shows magnified Sn-related RBS spectra. (b) P and Sn distribution in the Ge layer before and after FLA obtained by SIMS. FLA was performed for 3 ms at an energy density of 61.5 Jcm$^{-2}$.

Moreover, according to the RBS results the GeSn:P alloy quality is independent of the P fluence within the investigated range. The yield of the Ge signal in RBS/C is also low, confirming a high recrystallization efficiency during FLA. Since P atoms could not be resolved by RBS, SIMS was used as an alternative method, to observe the depth distribution of P in Ge before and after FLA. Fig. 1b shows the distribution of P (left axis) and Sn (right axis) in the GeSnP2 sample. The Ge layer containing Sn is homogenously doped with P and has a thickness of about 150 nm, while the total thickness of the n-type layer is 220 nm. The Sn distribution with an atomic concentration far exceeding the equilibrium solubility limit in Ge does not show a significant difference between the as-implanted stage and the flash lamp annealed one. A similar phenomenon is observed for P. The small variation in the P concentration between 100 and 170 nm can be caused by P trapping at the interface between GeSn and Ge. Nevertheless, the maximum P concentration and the depth distribution are almost unaffected by FLA. This confirms the RBS results as well as our assumption about the use of ms-range FLA for the realization of highly doped n-type GeSn:P layer with well controlled doping profile. Using r-FLA the implanted layer recrystallizes via ESPE. During ESPE the amorphous/crystalline interface moves towards the surface with a speed of about 10 m/s which is much faster than the diffusion of most dopants in Ge in the solid state [22]. Phenomena such as the snowplow effect are not observed in our annealed films [23, 24].

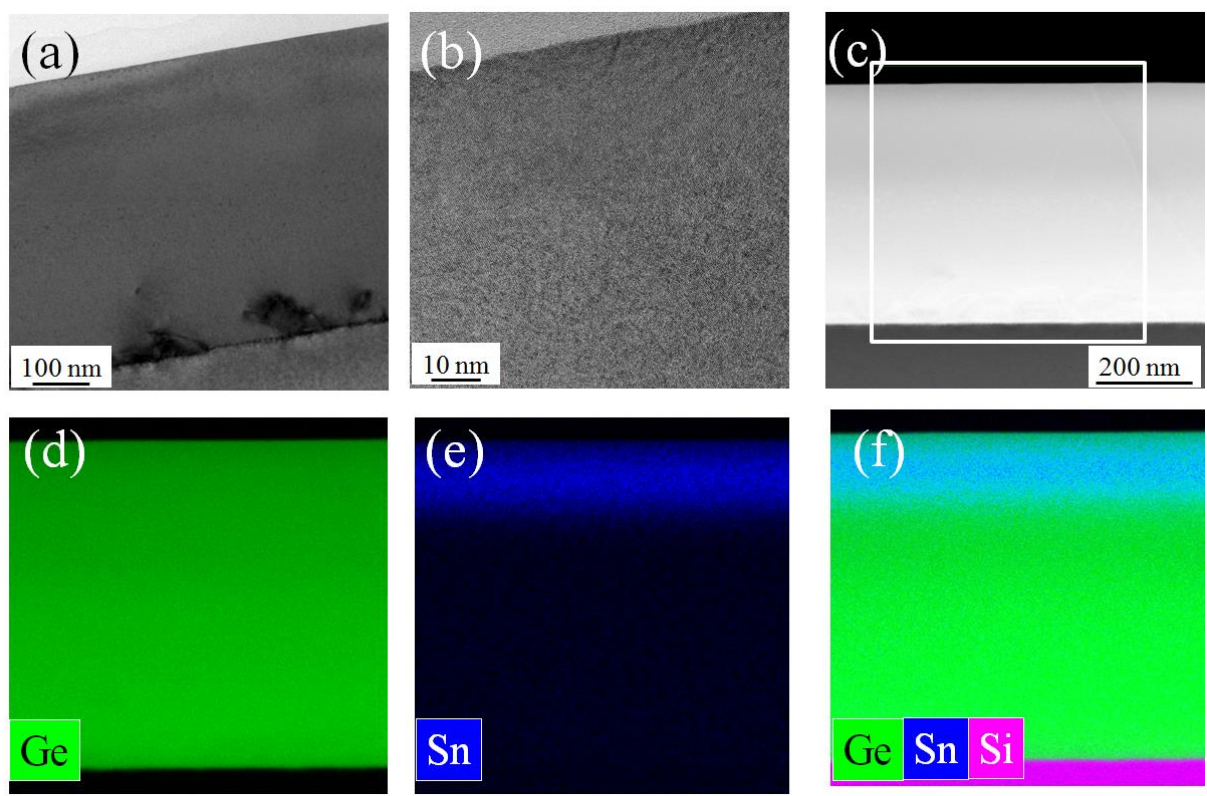

Figure 2. (a) Cross-sectional bright-field TEM micrograph and (b) high-resolution TEM image of Ge implanted with Sn and P ions after FLA for 3 ms. (c) Cross-sectional HAADF-STEM image as well as (d) Ge (green) distribution, (e) Sn (blue) distribution and (f) superimposed Si, Ge and Sn element distributions based on EDXS obtained in STEM mode for the region marked by the white rectangle in (c), respectively.

Figures 2 (a) and (b) show cross-sectional bright-field and high-resolution TEM images obtained from the GeSnP2 sample after FLA for 3 ms. Within the investigated area, the single-crystalline structure is restored, and even the end-of-range defects typical for ion-implanted semiconductors are not visible [25]. Only threading dislocations are detected at the interface between bulk Si and the grown Ge layer. The presence of such defects is typical for binary systems epitaxially grown with different lattice parameters [26]. In our case, the lattice mismatch between Ge and Si is in the range of 4 % and the threading dislocations annihilate completely within the first 200 nm of the grown layer. Figs. 2 (d-f) present the element distributions of Ge, Sn and Si based on EDXS spectrum imaging in scanning TEM mode from the window marked by the white rectangle in the HAADF-STEM image of Fig. 2c. As can be seen in Fig. 2 (e) and (f), Sn is evenly distributed in Ge over the implantation depth. No clusters or agglomerates of Sn were detected which is in good agreement with SIMS and RBS data.

### 3.2 Strain engineering
### 3.2.1 X-ray diffraction

In order to obtain the in-plane and out-of-plane strain distribution in the fabricated GeSn films XRD θ-2θ scans and reciprocal space maps were performed (see Fig. 3). The in-plane lattice parameter of the Ge virtual substrate is a=5.659Å while the out-of-plane lattice parameter is b=5.642Å. This means that already the as-deposited Ge layer is slightly biaxially tensile strained (ε=0.16%).

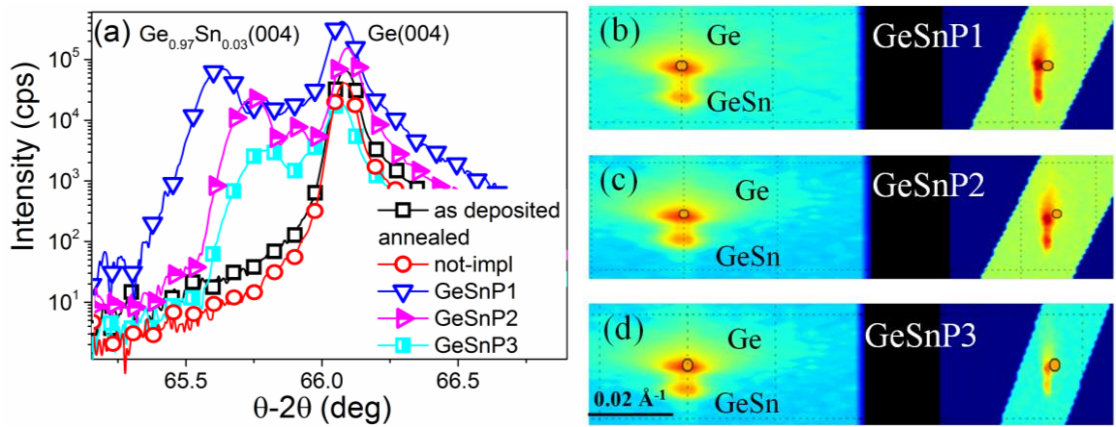

Figure 3. XRD θ-2θ scans around the Ge (004) diffraction of implanted and FLA treated samples, the virgin Ge-on-Si wafer, and the annealed sample without implantation (a) (curves are shifted vertically for clarity). (b), (c) and (d) show reciprocal space maps around the (004) (left panels) and (224) (right panels) reflections of Ge and GeSn after ion implantation and FLA. The orange circles in the images of reciprocal space maps describe the expected theoretical positions of a fully relaxed pseudomorphic Ge lattice.

The biaxial tensile strain obtained from XRD measurements is in good agreement with Raman data (see Fig. 4). The FLA performed on the as-deposited sample has no influence on the strain distribution, but the reciprocal space map suggests an improvement of the crystal quality due to the annealing of some defects, e.g., threading dislocations. The high-resolution θ-2θ XRD scan around Ge (004) reveals no significant changes in the Ge virtual-substrate due to ion implantation followed by non-equilibrium FLA, as well. After Sn implantation and annealing the additional peak at lower scattering angles describes the lattice expansion along [001]. The shift is caused by the formation of GeSn. Due to the fact that the lattice parameter of relaxed GeSn is bigger than that of the Ge virtual-substrate, the fabricated GeSn alloy is biaxially compressively strained. The XRD θ-2θ scan performed along (004) and reciprocal space maps show that the biaxial compressive strain in GeSn can be effectively compensated by co-doping with P. The in-plane lattice parameter "a" is the same in both the Ge virtual-substrate and the fabricated GeSn layer (a=5.659 Å) while the out-off plane lattice parameter "b" in GeSn alloys decreases from 5.671 Å in GeSnP1 down to 5.659 Å in GeSnP3.

Table 2. In-plane and out-off-plane lattice parameters of implanted and annealed samples. The calculated lattice parameter for relaxed GeSn$_{0.03}$ is also given.

| Sample | In-plane "a" (Å) | Out-off-plane "b" (Å) |
| --- | --- | --- |
| Ge pseudosubstrate | 5.659 | 5.642 |
| GeSnP1 | 5.659 | 5.671 |
| GeSnP2 | 5.659 | 5.665 |
| GeSnP3 | 5.659 | 5.659 |
| GeSn$_{0.03}$ | 5.683[*] | 5.683[*] |

[*] $a_{GeSn}$ in relaxed GeSn$_{0.03}$ was calculated using the following expression: $a_0 = a_{Sn} \times x + a_{Ge}(1-x) + bx(1-x)$, where $a_{Sn}$=6.4892Å, $a_{Ge}$=5.6579Å, $b$=0.00882Å and $x$=0.03 [27].

The values of the lattice parameters for different samples are summarized in table 2. Sn has a bigger covalent radius than Ge (140 pm vs. 120 pm, about 16% difference), which is why the lattice parameter of the relaxed GeSn alloy should be bigger than that of undoped Ge. The covalent radius of P (107 pm) is about 11% smaller than that of Ge. This means that P can efficiently reduce the lattice parameter of GeSn:P and compensate the biaxial compressive strain. Even the electronic contribution which normally expands the lattice cannot compensate the effect of P doping [28]. According to the theoretical prediction the lattice contraction in Ge doped with P to the concentration of $1.2\times10^{20}$ cm$^{-3}$ is in the range of 2 % [28]. In our case, the highest P concentration in the GeSn layer is close to $5\times10^{20}$ cm$^{-3}$ indicating a significant lattice contraction (close to 4 %) if the P lies in the substitutional position. Simultaneously, the biaxial tensile strain in the Ge virtual-substrate remains. If the Ge virtual-substrate will be replaced by the substrate with bigger lattice parameter e.g. Ge$_{100-x}$Sn$_x$ with x>3 a tensile strained ultra-doped GeSn layer can even be formed, favoring the formation of a direct band gap.

### 3.2.2 Micro-Raman spectroscopy

Figure 4 shows the micro-Raman spectra obtained from an as-deposited Ge-on-Si sample, a non-implanted but annealed sample, and samples double implanted with Sn and P followed by ms-range r-FLA. Due to the limited penetration depth of the green laser light in Ge and GeSn, the presented Raman spectra contain only the signal from the top 50 nm of the layer. The main peak observed from the as-deposited film corresponds to the transverse optical (TO) phonon mode in Ge, which is located at 299.9 ±0.1 cm$^{-1}$. In undoped and fully relaxed Ge the TO phonon mode should be located at 300.5 cm$^{-1}$. The shift of the TO phonon mode towards smaller wavenumbers can be explained in terms of tensile strain.

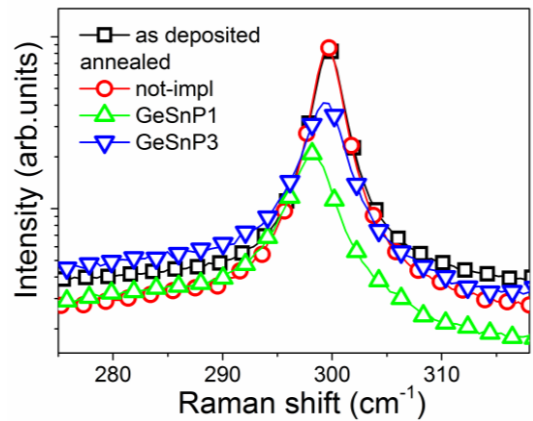

Figure 4. Micro-Raman spectra obtained from as-deposited and implanted Ge-on-Si wafers after r-FLA.

The lattice parameter of Si is 4.2% smaller than that of Ge, thus it is expected that the Ge layer grown on Si should be compressively strained. However, due to different thermal expansion coefficient between Ge and Si the Ge layer is tensile strained after the post-grown thermal treatment [29]. Moreover, the alloying of Ge with heavier elements like Sn or Pb shifts the position of the TO phonon mode towards shorter wavenumbers while the alloying with lighter elements like Si shifts the TO phonon mode towards longer wavenumbers.

Taking the shift of $\Delta\omega=-0.6 \pm 1$ cm$^{-1}$ into account, the strain can be calculated using the following expression [30]:

$$\varepsilon=[\Delta\omega/c]100\% \qquad (1)$$

where $\Delta\omega$ is the shift of the phonon mode position and *c* is a pre-factor (~ -300 for Ge). A positive $\varepsilon$ corresponds to the tensile strain while a negative $\varepsilon$ describes the compressive strain. According to the peak position of the TO phonon mode in the Raman spectrum, the as-grown Ge layer exhibits a tensile strain of 0.20±0.03 %, which is in good agreement with the value calculated from XRD data ($\varepsilon=0.16\%$). The r-FLA performed on the non-implanted sample does not affect the strain redistribution in the Ge layer.

In the case of relaxed GeSn alloy the Raman peak shift of the Ge-Ge LO mode is caused by the mass disorder (Ge vs. much heavier Sn) and the bond distortion (Ge-Ge bond vs. much longer Ge-Sn bond). Assuming a fully relaxed GeSn film with 3% of Sn the shift of the peak position of the TO phonon mode in the Raman spectrum in GeSn film ($\Delta\omega$) [30] can be calculated according to

$$\Delta\omega=\omega_{GeSn}-\omega_{Ge}=c\times x_{Sn} \qquad (2),$$

where the $\omega_{GeSn}$ and $\omega_{Ge}$ are the TO phonon modes in relaxed GeSn alloy and intrinsic Ge. $x_{Sn}$ is the Sn concentration in GeSn alloy and c=-82.8 after Ref. 30. Due to the lattice mismatch between the Ge virtual substrate and GeSn alloy is expected to be strained. The *in-plane* strain $\varepsilon$ can be also calculated from the Raman spectra for a given Sn content according to

$$\Delta\omega_{strain}=a\times x_{Sn}+b\times\varepsilon \qquad (3),$$

where $\varepsilon$ is the biaxial strain, a=-83.11 and b=-374.53, and $\Delta\omega_{strain}=$ is the shift of the TO phonon mode position with the respect to the relaxed GeSn [30]. The TO phonon mode in the sample GeSnP1 is located at 298.20 cm$^{-1}$. For the 3% of Sn in the relaxed GeSn alloy the peak position of the TO phonon mode in the Raman spectrum should be at 298.01 cm$^{-1}$. The shift of the peak position of the TO phonon mode in Ge$_{100-x}$Sn$_x$ for x<10% is caused by the alloy disorder and increases linearly with the Sn composition. According to formula (3) the GeSnP1 layer is compressively strained with $\varepsilon=-0.71$ %. The Raman spectrum obtained from GeSnP3 shows the TO phonon mode at 299.5 cm$^{-3}$ which indicates the existence of biaxial compressive strain of ε=-1.06 %. This is more than two times bigger than the strain calculated from the XRD data for sample GeSnP3 (ε=-0.42 %). The TO phonon mode position in Ge-based alloys depends on the element, strain and Fano effect [31]. The tensile and the compressive strains lead to a shift of the TO phonon mode towards shorter and longer wavenumbers, respectively. The Fano effect in heavily doped n-type Ge causes the shift of the TO phonon mode towards shorter wavenumbers. In our case, the shift of the TO phonon mode position towards longer wavenumbers with increasing P content in GeSn:P alloys can be explained by the change of the phonon energy due to the incorporation of P into GeSn and the generation of compressive strain. Therefore, only the strain calculated from the XRD data will be taken into account to analyze the change of the band gap as a function of strain. The calculation of strain from Raman spectra must be done carefully.

### 3.3 Band structure
### 3.3.1. Theoretical results

Sn incorporated into Ge modifies not only the lattice parameter, but, more importantly, the band structure of Ge [32, 33]. For a particular Sn concentration, it is possible to convert Ge from an indirect to a direct band gap material. It was shown that the amount of Sn needed to form a direct band gap in GeSn depends on the strain accumulated in the layer [34]. Therefore, the strain engineering will result in a band gap modification for a GeSn alloy with a fixed Sn concentration. The band gap of GeSn decreases when applying tensile strain and increases when the compressive strain is present. Moreover, it was shown that the band gap of Ge and GeSn alloys decreases with increasing the electron concentration due to the band gap renormalization [1, 35]. The band gap renormalization in relaxed GeSn alloy is in the range of 30 meV at an electron concentration of $2\times10^{19}$ cm$^{-3}$. At this doping level the Fermi level is still below the bottom of the Γ band and the Burstein-Moss effect can be neglected. Although for electron concentrations approaching the magnitude of $10^{20}$ cm$^{-3}$, the GeSn alloy becomes strongly degenerated with the Fermi level located inside the Γ band. Figure 5 shows the relative position of the Fermi energy ($E_F$) with respect to the conduction band minimum (CBM) as a function of electron concentration for a fully relaxed GeSn with 3.7 % of Sn. The inset in Fig. 5 shows the band structure of Ge$_{0.963}$Sn$_{0.037}$ with a horizontal line marking $E_F$ for an electron concentration of $1.1\times10^{20}$ cm$^{-3}$. The calculations were performed within the density functional theory (DFT) with the use of the FP-LAPW WIEN2k code [36]. A 54-atom supercell (3x3x3 multiplication of the primitive diamond unit cell) was used, allowing for studying the alloy in 1.85% of Sn intervals. Due to the size of the supercell the calculation of the band structure for GeSn with 3% of Sn was not possible but the difference of the band gap energy for the 3 and 3.7% of tin is only 20 meV. Within the supercell the Sn atom were distributed according to special quasirandom structures (SQS) to approximate a perfect alloy [37]. A full geometry optimization was performed with the use of the GGA-WC [38] functional and the electronic structure was calculated using the Tran and Blaha MBJLDA [39] functional to describe the band gap correctly.

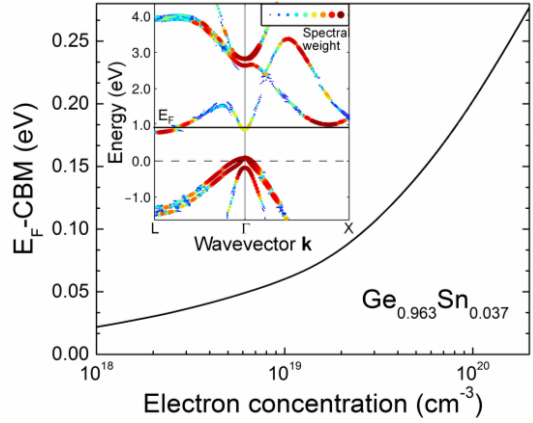

Figure 5. The calculated relative positon of the Fermi energy with respect to the conduction band minimum in $Ge_{0.963}Sn_{0.037}$ as a function of the doping level. The inset shows an unfolded band diagram of $Ge_{0.963}Sn_{0.037}$ with a horizontal line marking a position of the Fermi energy corresponding to an effective carrier concentration of $1.1 \times 10^{20}$ cm$^{-3}$. The color and the size of points correspond to the spectral weight of the eigenvalues.

The change of the Fermi energy as a function of electron concentration was obtained from an integrated density of states of $Ge_{0.963}Sn_{0.037}$, which was calculated on a dense mesh of 8x8x8 k-points (an equivalent of 24x24x24 mesh in a primitive unit cell) with the use of the modified tetrahedron method [40]. The unfolded band structure in the inset of Fig. 5 was obtained with the spectral weight approach as implemented in the fold2bloch [35] code, where the spectral weight of the eigenvalues is indicated by the color and size of the points. The presented simulation data do not take into account the band gap change due to the GeSn:P alloy formation with a reduced lattice parameter. The $\Delta E_F = E_F$-CBM in $Ge_{0.963}Sn_{0.037}$ was deduced to be 210 meV for a doping level of $1.1 \times 10^{20}$ cm$^{-3}$. Due to the low density of electron states at the Γ point of the Brillouin zone, $E_F$ is located deep in the conduction band and the effective absorption edge can be calculated as the energy difference between the valence band maximum and the $E_F$ position in the conduction band. Recently, Koerner *et al.* have shown that for electron concentrations lower than $5 \times 10^{18}$ cm$^{-3}$ and Sn concentrations below 6%, the band gap renormalization dominates and the Burstein-Moss Effect can be neglected [41]. It was also experimentally verified that the calculated band gap renormalization is in the range of 20 – 30 meV for doping levels up to $2\times10^{19}$ cm$^{-3}$ [35].

### 3.3.2. Experimental data

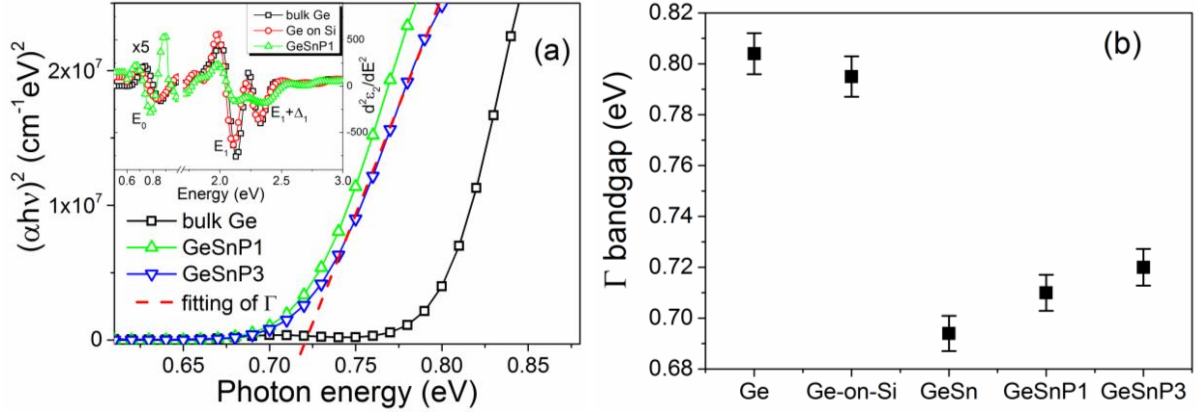

Figure 6. The plot of the $(\alpha h\nu)^2$ as a function of photon energy obtained from spectroscopic ellipsometry measurements (a) and the energy of the $\Gamma$ bandgap (b) in intrinsic bulk Ge, undoped tensile strained Ge-on-Si, undoped relaxed GeSn and heavily P doped GeSn samples with 3% of Sn extracted from (a) by linear fitting of the $(\alpha h\nu)^2$ vs. $h\nu$. The inset in (a) shows the second derivative of the imaginary part of the dielectric function ($d^2\varepsilon_2(E)/dE^2$) as obtained by spectroscopic ellipsometry from bulk Ge, non-implanted tensile strained Ge-on-Si and GeSnP1 samples.

To experimentally verify the band gap in GeSn alloy upon the doping and strain, the change of the energy of the interband transitions by spectroscopic ellipsometry was investigated. Figure 6 (a) shows the plot of the $(\alpha h\nu)^2$ vs. incident photon energy ($h\nu$) obtained from non-implanted and implanted samples followed by FLA. The $(\alpha h\nu)^2$ vs. $h\nu$ corresponds to the direct band gap absorption close to the band gap edge. The direct band gap of undoped and doped Ge samples was obtained by extrapolating the experimental data fitting to the photon energy axis. The undoped Ge and Ge-on-Si are relaxed and tensile strained, respectively. Therefore, the obtained value of the direct band gap corresponds to the transition between light holes (lh) and the $\Gamma$ valley [42]. The GeSnP alloys are compressively strained indicating that the extrapolated band gap is associated with the transition between heavy hole (hh) and the $\Gamma$ valley.

The inset in Fig. 6 (a) shows the low- and the high-energy part of the second derivative of the imaginary part of the dielectric function in the vicinity of the $E_0$, $E_1$ and $E_1$ +

$\Delta_1$ critical points obtained from bulk Ge, tensile strained virgin Ge-on-Si and GeSn:P alloy. The numerically obtained second derivatives were smoothed using Savitzky-Golay coefficients. The critical point $E_1$ in the GeSnP1 sample is red-shifted by 20 meV in comparison to virgin Ge. The critical point $E_1$ in GeSnP2 and GeSnP3 samples shifts back to the original position of the virgin Ge with increasing P concentration. The critical points $E_1$ and $E_1 + \Delta_1$ should not be affected by the Burstein-Moss effect. The observed red-shift is because the band gap renormalization takes place [43]. The $E_0$ critical point in GeSn alloys is significantly red-shifted in comparison with bulk Ge. In general, the red shift of $E_0$ with increasing P concentration is expected which would indicate a band gap narrowing in heavily doped n-type GeSn similarly to what happens in highly doped n-type Ge [44, 45]. However, the occurrence of the Burstein-Moss effect cannot be fully discarded for heavily doped GeSn layers. According to Hall effect measurements, the room-temperature effective electron concentration in P-doped samples is $7.2 \times 10^{19}$ cm$^{-3}$ for GeSnP1 and $10.5 \times 10^{19}$ cm$^{-3}$ for GeSnP3. The mean carrier concentrations for a particular doping level were estimated using the measured sheet Hall concentration and the thickness of the doped layer deduced from SIMS. The room-temperature electron mobility increases from 140 cm$^2$/Vs in GeSnP1 to 175 cm$^2$/Vs in GeSnP3. Since the Sn concentration is the same in all samples, the enhancement of the carrier mobility and the change of the band gap can be associated with the strain engineering and the band gap renormalization, respectively [46, 47].

Figure 6b shows the values for the direct band gap $\Gamma$ in bulk Ge, tensile stained Ge-on-Si, undoped GeSn and very heavily doped GeSn alloys. The final band gap of fabricated heavily doped GeSnP alloys is the result of the band gap renormalization, strain and Burstein-Moss effect. In undoped GeSn with 3% of Sn and compressive strain $\varepsilon = -0.42\%$, $E_g(\Gamma)$ is around 0.75 eV. Assuming the band gap renormalization at the electron concentration of $10.5 \times 10^{19}$ cm$^{-3}$ for GeSnP3, the band gap should decrease by $\Delta E = 110$ meV to $E_g(\Gamma) = 0.64$ eV. However, the measured direct band gap in GeSnP3 was found to be $E_g(\Gamma) = 0.72$ eV which is 80 meV bigger than expected. The increase of the band gap in GeSn:P alloy with increasing effective carrier concentration above $7 \times 10^{19}$ cm$^{-3}$ must be caused by the Burstein–Moss effect due to ultra-high n-type doping [48]. Therefore, in contrast to intrinsic or lightly doped n-type GeSn, in strongly degenerated n-type GeSn a blue-shift of the effective absorption edge is expected. This means that even strain-free GeSn with the Sn content in the range of 3% and $E_F$ located deep in the direct band gap may become a quasi-direct band gap semiconductor (see the inset in Fig. 5).

## 4. Conclusions

Highly-doped n$^{++}$ GeSn:P-on-Si alloys have been fabricated using ion beam implantation and post-implantation non-equilibrium thermal processing. Effective strain and Fermi energy engineering have been achieved by P co-doping and r-FLA. The compensation of biaxial compressive strain and the lattice parameter reduction results in a band gap change accompanied by an enhancement of the carrier mobility. Both the reduction of the lattice parameter and the Burstein-Moss effect have been proven to be responsible for the blue shift of the absorption edge and partial compensation of the band gap renormalization in GeSn:P alloys. The whole fabrication procedure is fully compatible to CMOS technology, which paves the way for Ge-based optoelectronics.

**Acknowledgement**

Support by the Ion Beam Center (IBC) at HZDR is gratefully acknowledged. We would like to thank Andrea Scholz for XRD measurements and Romy Aniol for TEM specimen preparation. This work was partially supported by the German Academic Exchange Service (DAAD, Project-ID:57216326) and National Science Centre, Poland, under Grant No. 2016/23/B/ST7/03451. Y. Berencén would like to thank the Alexander-von-Humboldt foundation for providing a postdoctoral fellowship. The funding of TEM Talos by the German Federal Ministry of Education of Research (BMBF), Grant No. 03SF0451 in the framework of HEMCP is gratefully acknowledged.